\documentclass{kapproc} 

\usepackage{t1enc}

\usepackage{procps} 

\usepackage[dvips]{graphicx}

\upperandlowercase

\kluwerbib 

\begin{document}

\articletitle{K-luminous galaxies at z $\sim2$}

\articlesubtitle{Metallicity and B Stars}

\author{Duilia de Mello\altaffilmark{1,2} and the K20 Team}
\altaffiltext{1}{Laboratory for Astronomy and Solar Physics GSFC/NASA - Catholic University of America}
\altaffiltext{2}{Johns Hopkins University, Baltimore, MD21218, USA}
\email{duilia@ipanema.gsfc.nasa.gov}

\begin{abstract}
We present the results from the analysis of the composite
spectrum of five near-IR luminous ($K < 20$) galaxies at $z\sim2$. 
Several of the strongest absorption lines are present in 
the merging galaxy NGC 6090 UV spectrum and are not present in the spectra of 
Lyman Break Galaxies at $z\sim3$. They were identified as SiIII~1296, CIII~1428, SiII~1485, 
and Fe~$\sim$1380 \AA\ which are photospheric lines typical of B stars. 
A metallicity higher than solar is suggested by comparing 
the pure photospheric lines known as the 1425\AA\ index (SiIII, CIII, FeV) 
with Starburst99 models. The evidence of high metallicity, together with the 
high masses, high star-formation rates, and possibly strong clustering, 
suggest that these galaxies are candidates to become massive 
spheroids.

\end{abstract}


\section{Introduction - The K20 Survey}

A key open question in galaxy evolution is the epoch of formation of
massive spheroidal galaxies. As the rest-frame optical--near-infrared
traces the galaxy mass, the $K_s$-band allows a fair selection of
galaxies according to their masses up to $z\sim$2. Based on this, a VLT 
spectroscopic survey of about 500 galaxies with $K_s<20$ in the GOODS southern field was conducted 
(Cimatti et al. 2002). In this contribution (see also de Mello et al. 2004), we analyze the average spectrum of five K20
galaxies at $1.7<z<2.3$ with the highest S/N ratio among the 
ones presented in Daddi et al. (2004). These K$<20$ galaxies at $z\sim2$ appear to be massive
($\ge$ 10$^{11}$M$_{\odot}$) and have high star-formation rates (SFR 100--500 M$_{\odot}$/yr) 
(Daddi et al. 2004), thus qualifying as good candidates for assembling/forming massive 
early-type galaxies.

\begin{figure}[ht]
\begin{minipage}{157pt}
\center{
\includegraphics[width=157pt,draft=false,]{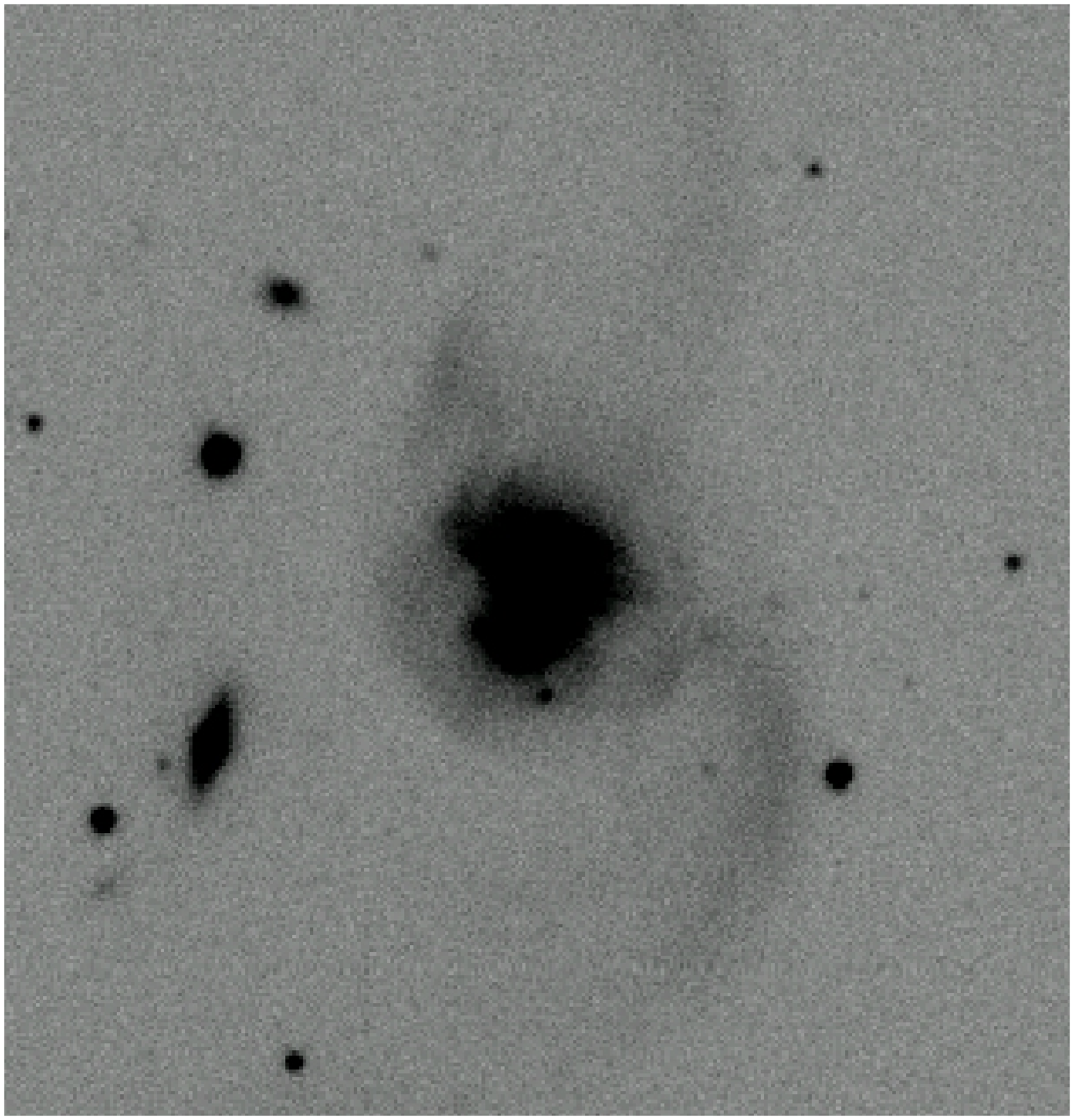}
\caption[]{\label{f1} Sloan r-band image of NGC~6090 extracted from
the National Virtual Observatory ($\sim$4 arcmin$^{2}$).
}}
\end{minipage}
\begin{minipage}{187pt}
\center{
\includegraphics[width=187pt,draft=false,]{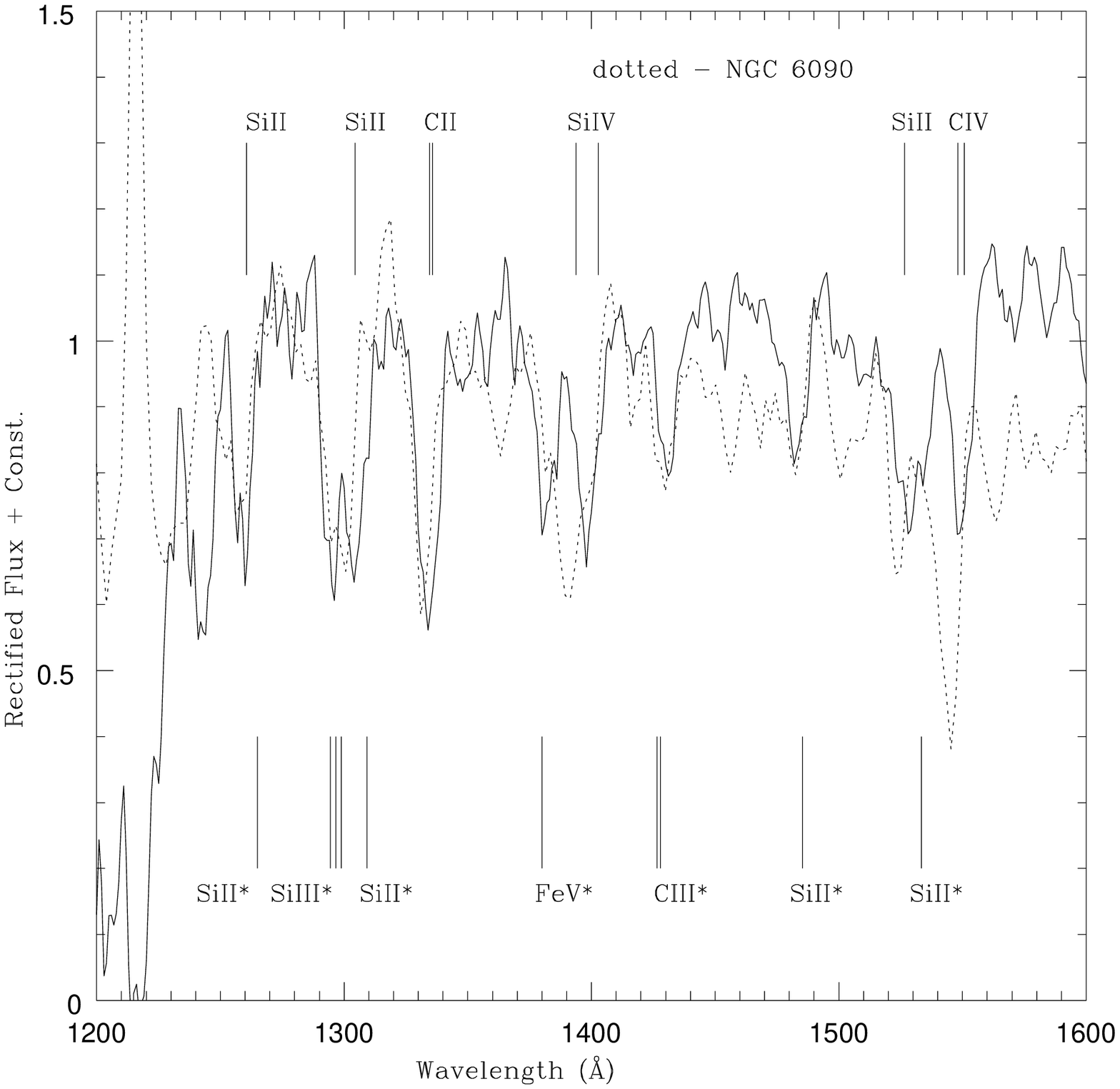}
\caption[]{\label{f2} The spectrum of NGC~6090 (dotted lines) and the K20
  composite spectrum. 
Photospheric lines are marked with $\ast$.}}
\end{minipage} 
\end{figure}

\section{Local starbursts and B stars}

We have compared the K20 average spectrum with the local starburst 
galaxies NGC 1705, NGC~1741, NGC~4214, and NGC~6090. 
The best match is obtained with NGC~6090 which is an interacting system at 
v $\sim$ 9062 km/s in the process of merging. It is a luminous infrared galaxy 
(log L$_{\rm IR}$ = 11.51; Scoville et al. 2000) with a number of luminous clusters triggered 
by the galaxy-galaxy interaction. The Sloan Digital Sky Survey r-band image (Fig.\ref{f1})   
\footnote{We acknowledge use of the National Virtual Observatory, which is funded by
the National Science Foundation under Cooperative Agreeement AST0122449 with
The Johns Hopkins University.} shows tidal tails that extend several arcminutes
from the two merging objects. NICMOS/HST images of NGC~6090 (Scoville et al. 2000) show the 
inner site of the interaction in more detail, where a less massive galaxy seems to be 
merging with a disk. The spectrum of NGC~6090 taken by the HUT during the Astro-2 mission 
(Gonzalez Delgado et al. 1998) is shown in Fig.\ref{f2}. It has several absorption lines which are 
similar in strength to the K20 composite spectrum, such as the photospheric lines SiIII $1295$ \AA, 
CIII $1430$ \AA\ and SiII $1485$ \AA\ and a marginally weaker FeV $1380$ \AA\ line.

\begin{figure}[ht]
\begin{minipage}{170pt}
\center{
\includegraphics[width=170pt,draft=false,]{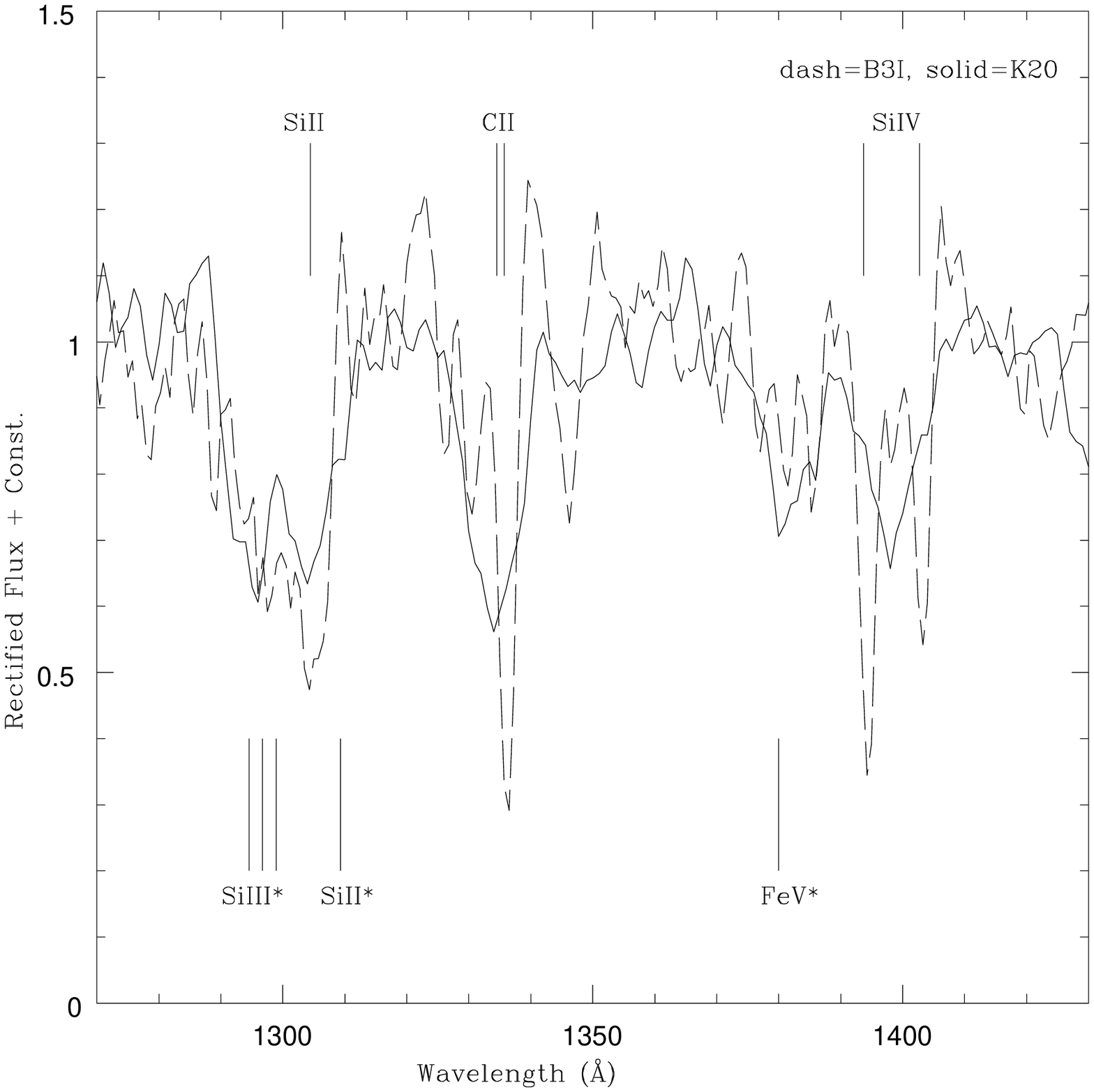}
\caption[]{\label{f3} The spectrum of a B3I star (dashed) and the K20 composite spectrum.}}
\end{minipage}
\begin{minipage}{170pt}
\center{
\includegraphics[width=170pt,draft=false,]{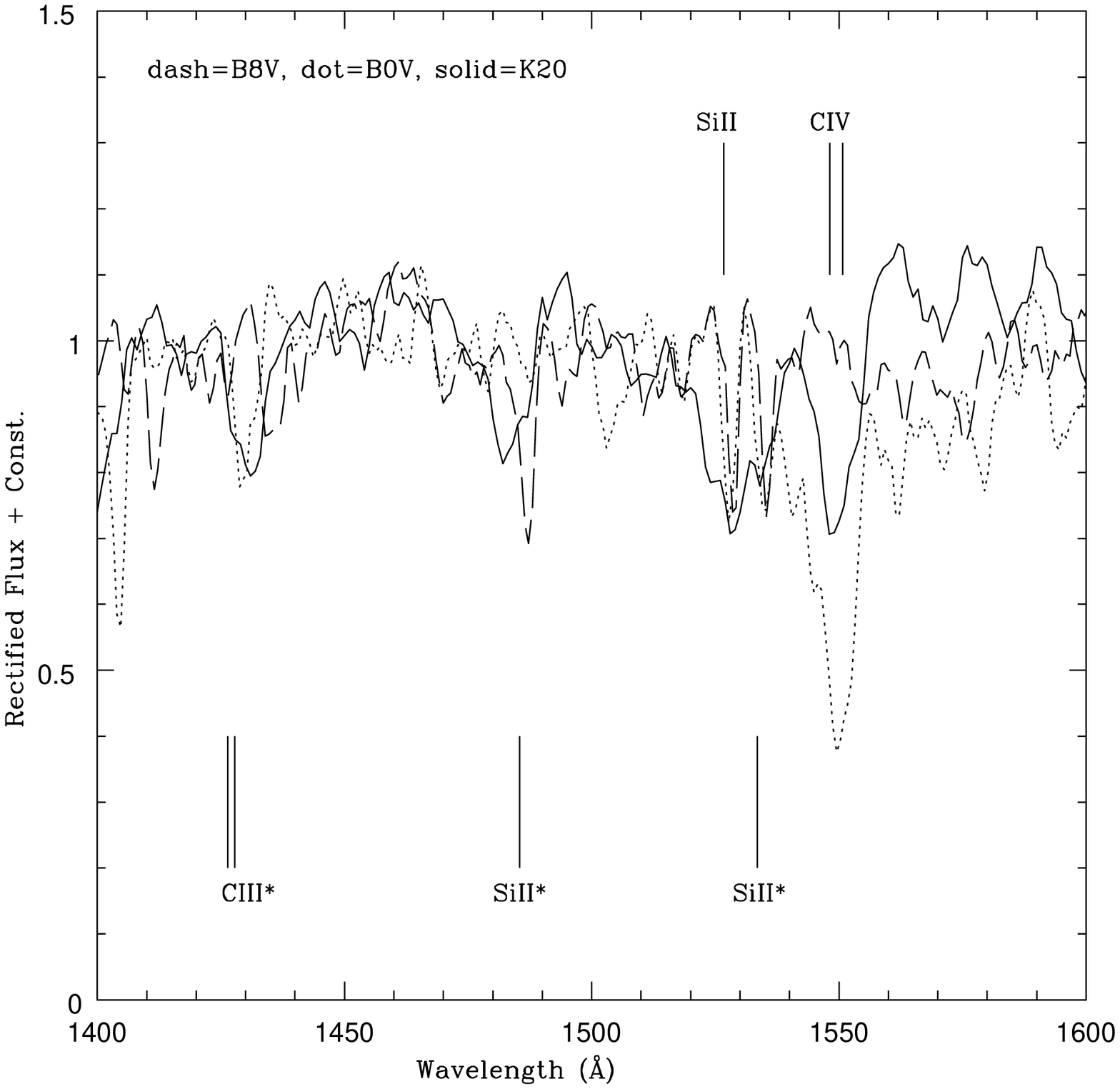}
\caption[]{\label{f4} The spectrum of a B8V (dashed) star and a B0V (dotted) star and the K20 composite
spectrum. Photospheric lines are marked with $\ast$.}}
\end{minipage}
\end{figure}

In order to search for the stellar population 
which contains these photospheric lines, 
we have examined a far-UV (IUE) library of Milky Way 
OB stars (de Mello et al. 2000). The similarity
between the spectra of B stars and the K20 composite spectrum is remarkable.
In Fig.\ref{f3} and Fig.\ref{f4} we show the average spectrum of two main sequence stars (B0V and B8V) 
and a supergiant (B3I) where the main photospheric lines are
identified. B stars live longer than the more luminous short-lived O stars and become 
a major source of the UV flux in the integrated spectrum of starbursts. 
The photospheric lines found in the spectrum of K20 galaxies and in NGC~6090 
are stellar features of B stars. 

\section{Lyman Break Galaxies and Metallicity}

\begin{figure}[ht]
\begin{minipage}{177pt}
\center{
\includegraphics[width=177pt,draft=false,]{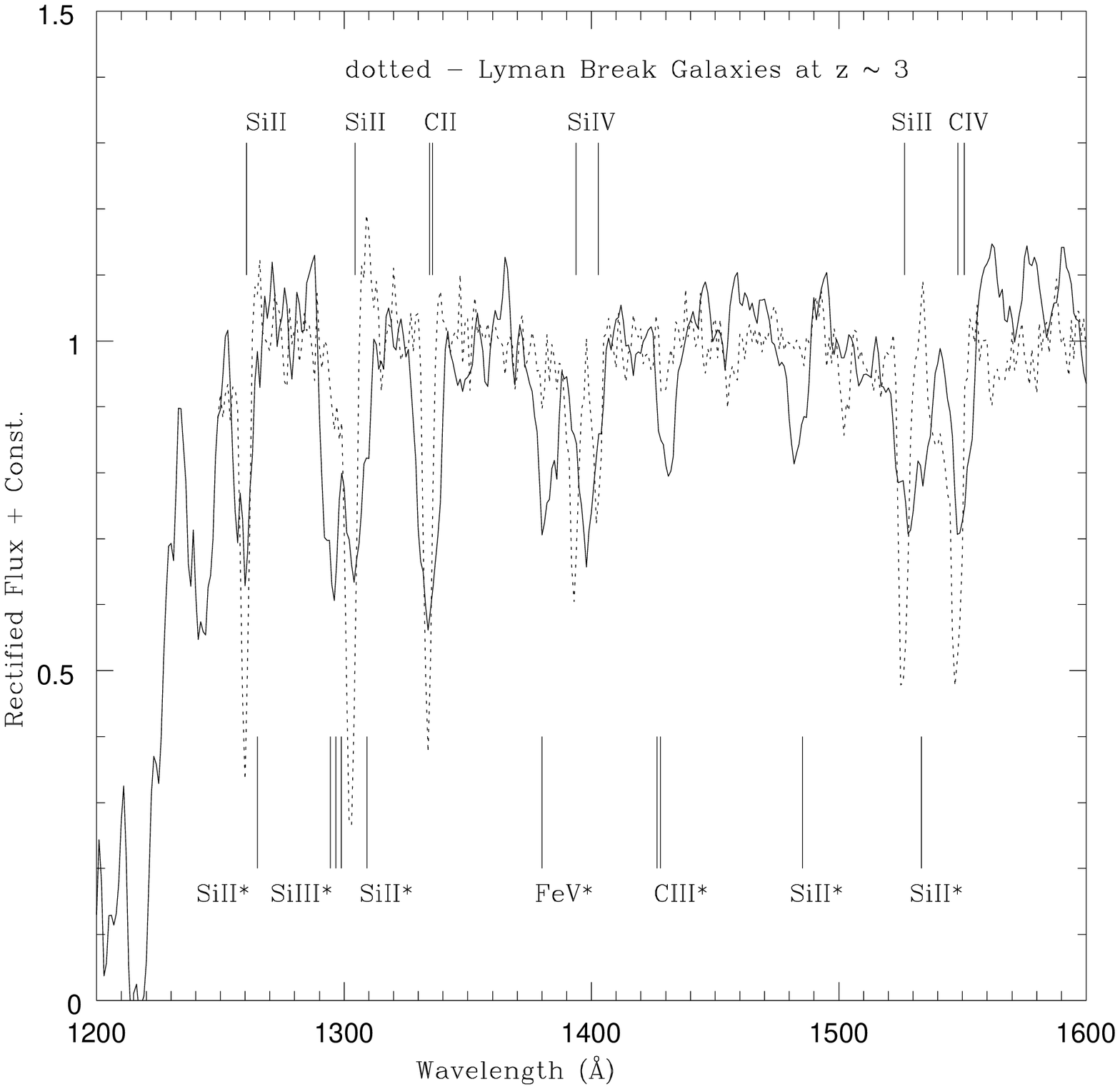}
\caption[]{\label{f5} The spectrum of Lyman Break Galaxies (dotted) and the K20 composite spectrum.}}
\end{minipage}
\begin{minipage}{140pt}
\center{
\includegraphics[width=140pt,draft=false,]{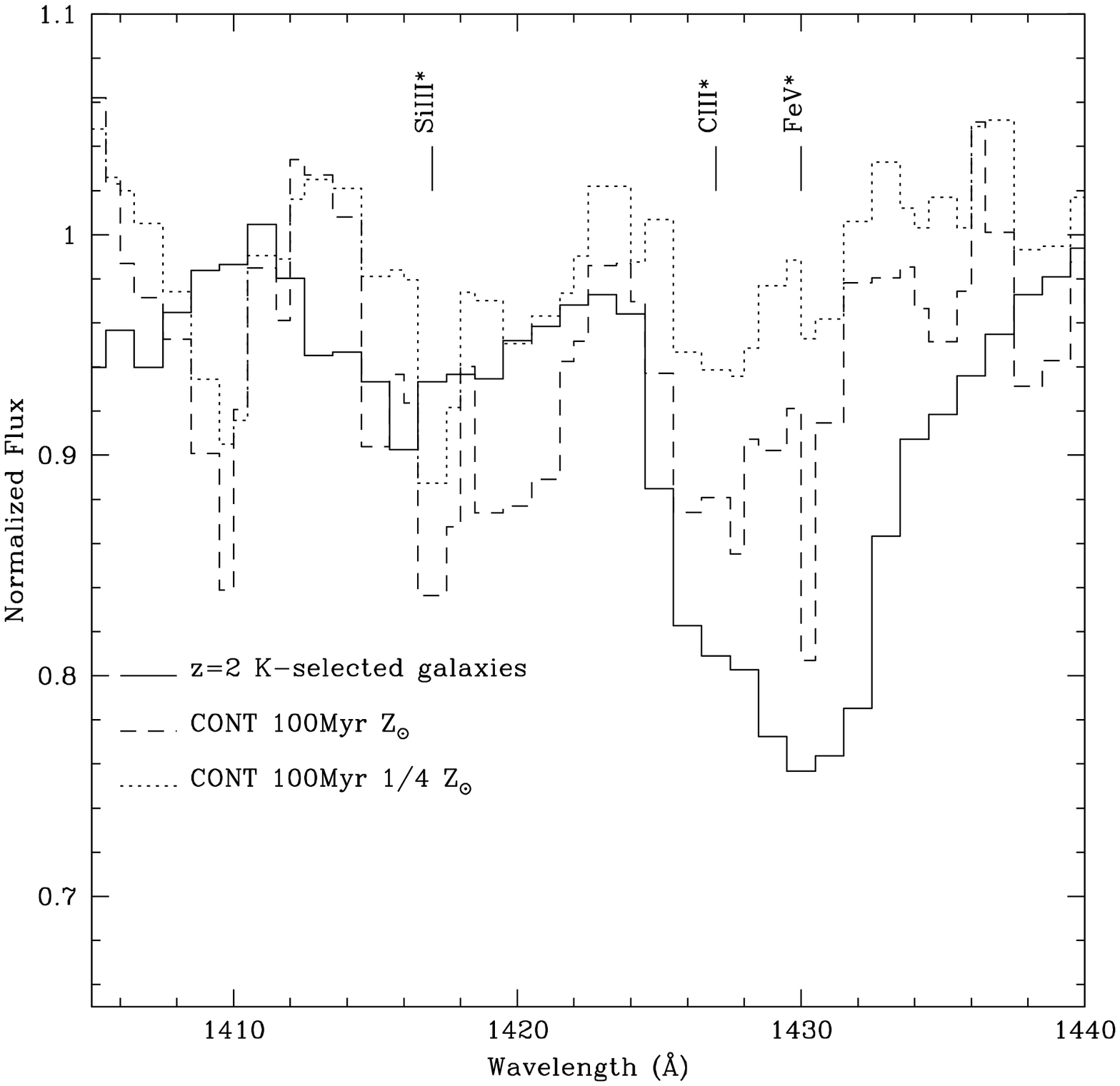}
\caption[]{\label{f6} The 1425\AA\ index region of the K20 composite spectrum . Starburst99 models for solar metallicity (dashed) and  
  0.25 Z$_{\odot}$ (dotted).}}
\end{minipage}
\end{figure}

We have also compared the K20 average spectrum with the average Lyman Break Galaxies (LBGs)
spectrum (Shapley et al. 2003) at z$\sim$3. The best match is obtained with the
LBGs without Lyman-$\alpha$ emission (Fig.\ref{f5}). The most striking differences between 
the LBG composite spectra and the K20 average spectrum are the photospheric lines described 
above. One caveat that one has to have in mind, before further intrepreting
this comparison, is the fact that the LBGs' average-spectrum contains several hundred spectra,
whereas the spectrum we are presenting here is the average of only 5 objects. Therefore, 
a few peculiar objects could be present and co-addition of a larger number of spectra is 
desirable for the future in order to smooth out the contribution of individual objects.
Nevertheless, they comprise an interesting class of objects that might be
important in the galaxy evolution scenario. 

We have used the  pure photospheric lines known as 
the 1425\AA\ index (SiIII, CIII, FeV) (Leitherer et al. 01) 
to estimate metallicity (Fig.\ref{f6}), since it does not strongly depend on age. The equivalent 
width of the index is 2.3$\pm$0.4 \AA, a value much larger than in Starburst99 models (Leitherer et al. 1999)
for solar (1.5\AA) and LMC metallicity (0.4\AA). It corresponds to  
models with continuous star formation, fewer massive stars and metallicities 
larger than solar (Fig. 11 in Rix et al. 2004). Recently, Shapley et al. (2004) also suggested
that K$<20$ galaxies at $2.1<z<2.5$ have at least solar metallicity from near-IR spectroscopic measurements 
of seven galaxies. Near-IR spectroscopy of a larger sample of K20 galaxies is needed 
in order to confirm the metallicities and estimate the relative importance of O and B stars in these 
objects. 

\end{document}